\documentclass[aps,twocolumn,a4paper,showpacs]{revtex4}
\usepackage{graphicx}
\usepackage{amsmath}
\usepackage{amssymb}
\usepackage{enumerate}
\usepackage{subfigure}
\usepackage{tabularx}
\newcommand{\be}{\begin{equation}}
\newcommand{\ee}{\end{equation}}
\newcommand{\ben}{\begin{eqnarray}}
\newcommand{\een}{\end{eqnarray}}
\newcommand{\bes}{\begin{subequations}}
\newcommand{\ees}{\end{subequations}}

\newcommand{\bb}{\bibitem}

\begin{document}
\title{Topological strength of magnetic skyrmions}
\author{D. Bazeia, J.G.G.S. Ramos, and E.I.B. Rodrigues}
\affiliation{Departamento de F\'\i sica, Universidade Federal da Para\'\i ba, 58051-900 Jo\~ao Pessoa, PB, Brazil}
\begin{abstract}
This work deals with magnetic structures that attain integer and half-integer skyrmion numbers. We model and solve the problem analytically, and show how the solutions appear in materials that engender distinct, very specific physical properties, and use them to describe their topological features. In particular, we found a way to model skyrmion with a large transition region correlated with the presence of a two-peak skyrmion number density. Moreover, we run into the issue concerning the topological strength of a vortex-like structure and suggest an experimental realization, important to decide how to modify and measure the topological strength of the magnetic structure.   
\end{abstract}
\date{\today}
\maketitle

\section{Introduction}

Topological structures play important role in nonlinear science and may come out as kinks, domain walls, vortices, strings, monopoles, skyrmions and other localized solutions \cite{v,hs,ms}. They are static solutions that appear in different spatial dimensions, and in this work we study planar systems, focusing attention on magnetic domains \cite{hs} that behave as magnetic spin textures of the vortex or skyrmion type. Such magnetic excitations are of current interest, and have been investigated in a diversity of contexts, in particular in the recent works \cite{s1,s2,s3,s4,s5,s5a,s6,s7,s8,s9,s10,s10a,s11,s12x,s13,s14,s14x,s14y,s15,s16,s16x,s17,ezawa}, where fabrication and tailoring of skyrmions and skyrmion lattices are important steps toward applications at the nanometric scale, where such spin textures are conceived.

Since the magnetic structures that we study are marked by topological features that describe their skyrmion numbers, we concentrate on issues related to the topology and stability that the localized structures may engender. This is a topic of current interest and we follow the lines of \cite{s11,s12x,s13,s14,s14x}, which deals with the possibility to control, enhance and measure the strength the topology induces into the solutions. To do this, we follow the recent work \cite{s17}, in which we have studied vortex and skyrmion properties starting from an exactly solvable model described by a scalar field in two spatial dimensions, as introduced in \cite{bmm}.

We focus attention on the recent investigations \cite{s12x,s13,s14x}, where the authors study skyrmions and discuss the measurement \cite{s12x}, and splitting and enhancement \cite{s13,s14x} of such topological structures. In particular, in \cite{s12x} the authors construct interesting experimental samples, with a vortex on top of a magnetic material, with the magnetic material having two distinct configurations, one with the magnetic spins pointing upward in the out-of plane direction, and the other with the spins pointing downward. They then apply external in-plane magnetic field, which they control and vary, and show that as the external field increases, one of the skyrmions is destroyed before the other one, indicating the strength the topology plays in the magnetic structures. However, we think that the skyrmions created in \cite{s12x} are in fact vortices of a specific magnetic material (they used Co circular disk) which sit on top of another material (they used Ni film, grown epitaxially on a Cu(001) substrate) with the Ni spins pointing up and down in the out-of-plane direction, and the effect measured shows how the up and down out-of-plane magnetic contributions of the Ni film change the topological strength of the vortex structure which sits on top of it.

In refs.~\cite{s13,s14x} the authors investigate several issues, among them the two-peak appearance in the topological charge density, which is correlated with the presence of a large {\it transition region}, representing the {\it internal structure} of the topological skyrmion. This is another issue that we study in this work, with the investigation of a model which appeared before in \cite{bmm}. The motivation here is that in the one-dimensional case studied in \cite{bmm}, one found a domain wall similar to the magnetic domain wall that appeared experimentally in ${\rm Fe}_{20}{\rm Ni}_{80}$ thin film, in a constrained geometry \cite{is}. As we are going to show, the model can be used to map skyrmion with unit skyrmion number, but with the number density having a two-peak formation, correlated with the presence of a large transition region, the internal structure of the magnetic solution. In the current work, the effect is related to the fractional self-interactions that describe the model, and in Ref.~\cite{s14x} it is supposed to appear in consequence of the Rashba spin-orbit coupling. We recall that the Rashba spin-orbit coupling has been studied in several works, in particular in the experimental and theoretical contexts in Refs.~\cite{R1,R2,R3,R4,R5}.

With these motivations in mind, in the current work we describe how to construct theoretically, structures having skyrmion number $1$, $1/2$, and $0$, and show that they all crucially depend on specific physical properties of the material under investigation, so they have to be generated by specific materials, each one with its specific features. This leads us to suggest that the presence of a magnetic vortex on top of an out-of-plane aligned spin arrangement of another magnetic material is still a vortex, with skyrmion number $1/2$. It may be a hybrid structure, and to describe our point of view, we organize the work as follows. In the next section we introduce the general framework and briefly review the construction of skyrmions with topological numbers $1$ and $1/2$, and then study a new model, which may induce splitting of the localized structure. We go on and construct other localized structures with vanishing skyrmion number in Sec.~\ref{nt}. Next, in Sec.~\ref{sta} we investigate the stability of the new localized solutions presented in the work. In Sec.~\ref{td} one uses the topological charge density to show how it behaves for the several distinct structures constructed in Secs.~\ref{remarks} and \ref{nt}, and in Sec.~\ref{top} we deal with the topology of these localized structures. We end the work in Sec.~\ref{end}, where we include our comments and conclusions.

\section{Framework}
\label{remarks}

We start reviewing the main results of the work \cite{s17} on the subject. One supposes that the magnetic material is homogeneous along the ${\hat z}$ direction, such that the magnetization ${\bf M}$ is a vector with unit modulus that depends only on the planar coordinates, such that ${\bf M}={\bf M} (x,y)$ and ${\bf M}\cdot{\bf M}=1$. 

To describe skyrmions, one uses the magnetization ${\bf M}$ to introduce the skyrmion number, which is a conserved topological quantity, defined by
\be\label{Q}
Q=\frac{1}{4\pi}\int_{-\infty}^{\infty}\!\!\! dx\, dy \;{\bf M}\cdot\partial_x{\bf M}\times\partial_y{\bf M}.
\ee
In this work we concentrate on planar systems and focus on the case of helical excitations, with ${\bf M}={\bf M}(r)$, which only depends on the radial coordinate but is orthogonal to the radial direction, that is, ${\bf M}\cdot{\hat r}=0$, in cylindrical coordinates. We then write the magnetization as
\be\label{M}
{\bf M}(r)={\hat \theta}\cos\Theta(r) + {\hat z}\sin\Theta(r),
\ee 
where $\Theta(r)$ is the single degree of freedom which we use to describe the magnetic excitations. We can use this magnetization to see that the skyrmion number is given by, after changing variables from $(x,y)$ to $(r,\theta)$,
\be
Q=-\frac12\sin\Theta(\infty)+\frac12\sin\Theta(0).
\ee
The topological profile of the solution $\Theta(r)$ is then related to its value at the origin, and the asymptotic behavior for larger and larger values of $r$.

To model skyrmion solutions analytically, we take advantage of the recent study \cite{s17} and consider
\be\label{T}
\Theta(r)=\frac{\pi}2\phi(r)+\delta,
\ee
where $\delta$ is a constant phase, which is used to control the magnetization at the center of the magnetic structure, determined by $r=0$. Also, we suppose that the scalar field $\phi$ is homogeneous and dimensionless quantity which is described by the planar system investigated in \cite{bmm}. The Lagrange density ${\cal L}$ that controls the scalar field $\phi$ has the form
\be\label{model}
{\cal L}=\frac12\dot\phi^2-\frac12 \nabla\phi\cdot\nabla\phi-U(\phi),
\ee
where dot represents time derivative, and $\nabla$ is the planar gradient. We search for time independent and spherically symmetric configuration, $\phi=\phi(r)$, and consider $U=U(r,\phi)$ in the form
\be\label{poten}
U(r,\phi)=\frac1{2r^2}P(\phi),
\ee
with $P(\phi)$ an even polynomial, which is supposed to contain non-gradient terms in $\phi$.

An important motivation to take \eqref{model} and \eqref{poten} to describe the scalar degree of freedom is that it provides analytical solutions that can be used to improve our understanding on the topological behavior of such structures. Another motivation comes from the fact that in the one-dimensional case studied before in \cite{bmm} with a potential dependent on an odd-integer parameter (which we shall further explore below), one found a domain wall behavior that nicely describes the magnetic domain wall that appeared experimentally in a ${\rm Fe}_{20}{\rm Ni}_{80}$ thin film element, in constrained geometries at the nanometric scale \cite{is}. 

We then go on and use \eqref{model} and \eqref{poten}, to get the field equation
\be
r^2\frac{d^2\phi}{dr^2}+ r \frac{d\phi}{dr}-\frac12\frac{dP}{d\phi}=0.
\ee
Here, the energy for static solution $\phi(r)$ is given by
\be\label{Ea}
E=2\pi\int_0^\infty rdr\rho(r),
\ee
with $\rho(r)$ being the energy density, such that
\be\label{Eb}
\rho(r) = \frac12 \left(\frac{d\phi}{dr}\right)^2+ \frac{1}{2r^2}P(\phi).
\ee 

We note from the above discussion that a specific model is defined when we specify the polynomial $P(\phi)$. So, we follow as in the Landau expansion for non-gradient contributions, to suggest and describe distinct models using different expansions for $P(\phi)$. This allows that we investigate topological and non-topological structures and illustrate the general situation with different models in the next subsections.

\begin{figure}[t!]
\centerline{\includegraphics[scale=0.38]{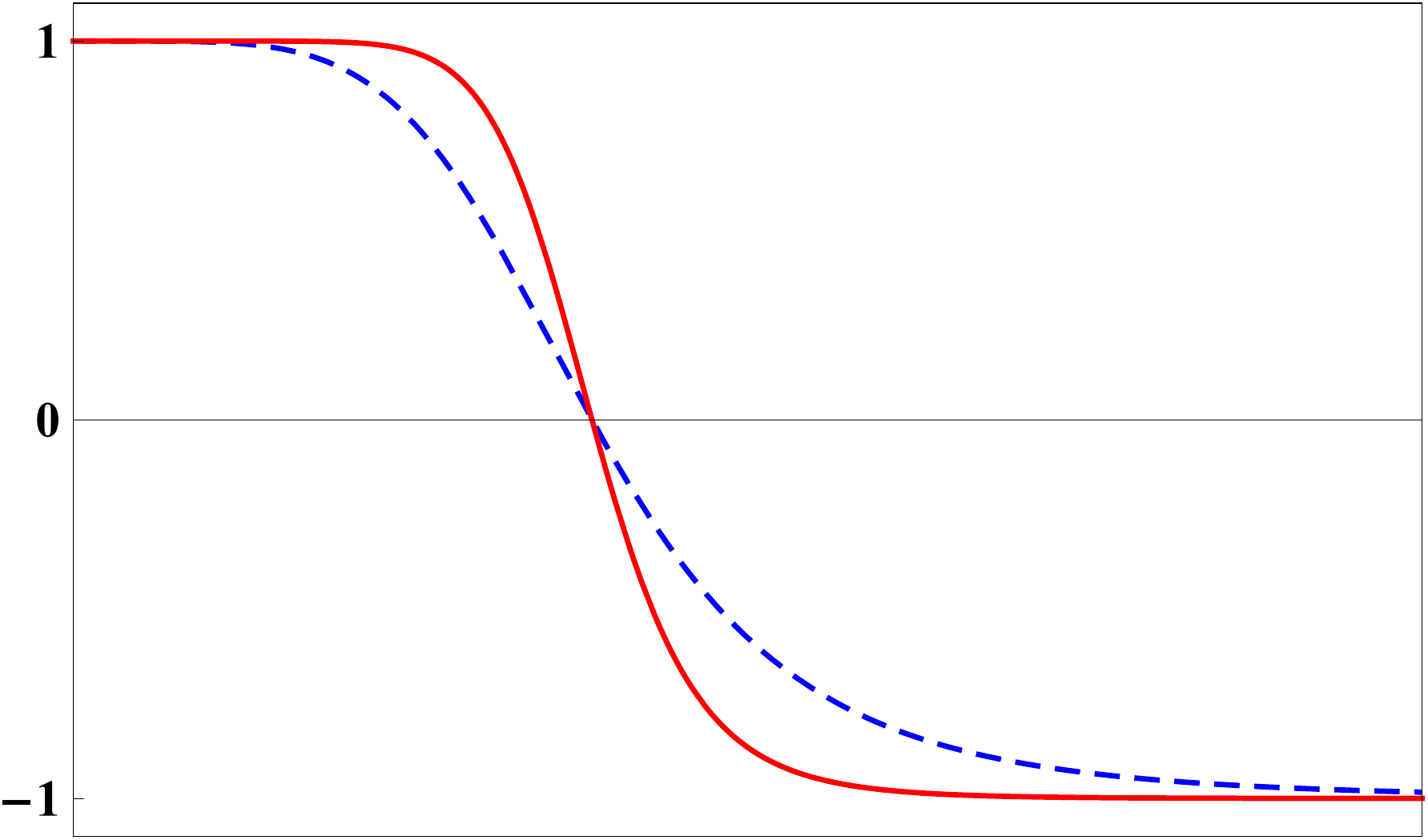}}
\caption{(Color online) The solution $\phi(r)$ for the model \eqref{p4}, depicted for $s=0.6$ and for $s=0.8$, with dashed/blue and solid/red lines, respectively.}\label{fig1}
\end{figure}
\begin{figure}[t!]
\centerline{\includegraphics[scale=0.38]{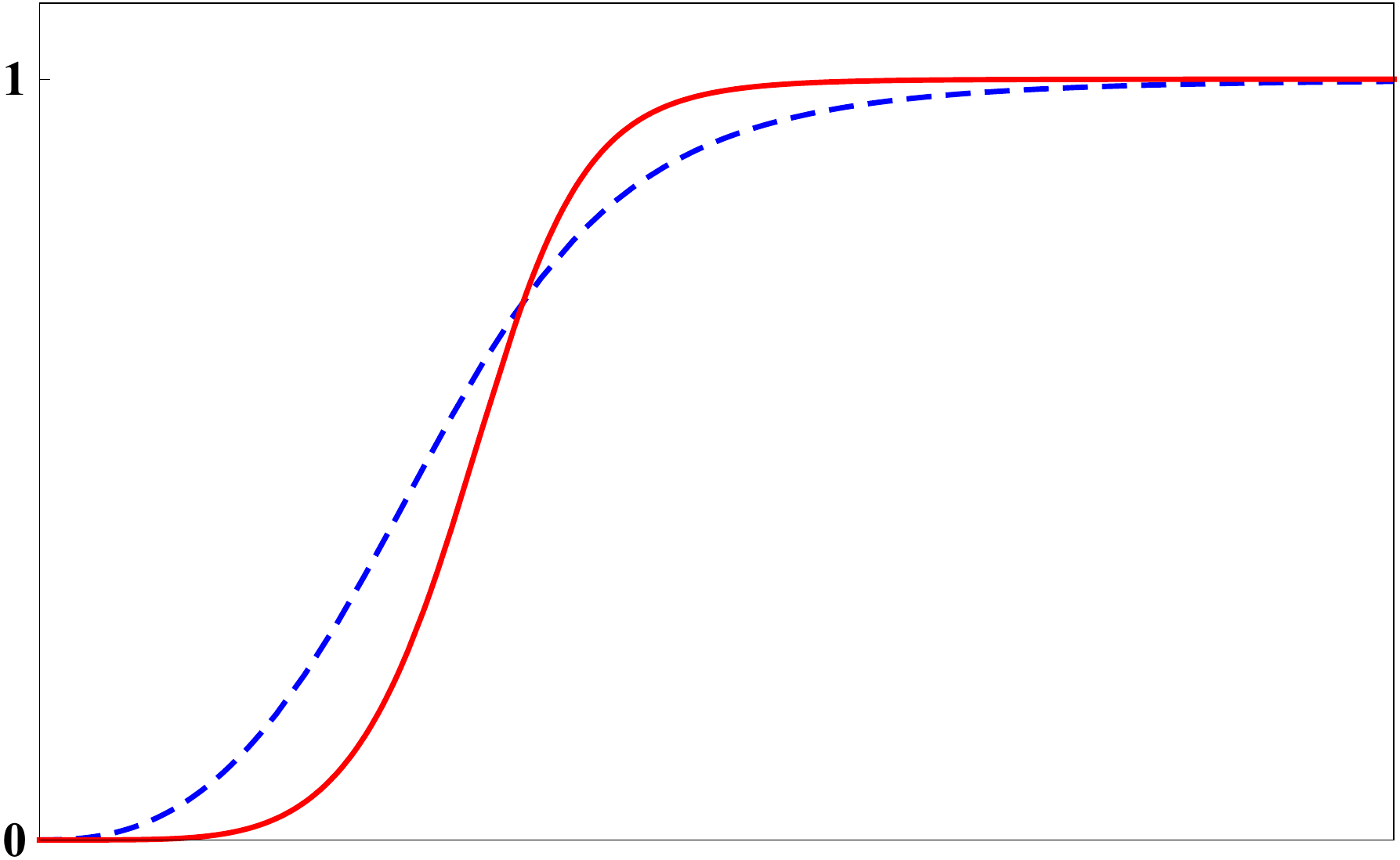}}
\caption{(Color online) The solution $\phi(r)$ for the model \eqref{p6}, depicted for $s=0.6$ and for $s=0.8$, with dashed/blue and solid/red lines, respectively.}\label{fig2}
\end{figure}

\subsection{Skyrmion} 

Let us consider a spin texture with skyrmion number $Q=1$. This case is described by the model with polynomial $P(\phi)$ given by
\be\label{p4}
P(\phi)=\frac1{(1-s)^2}(1-\phi^2)^2,
\ee
where $s$ is a real parameter, $s\in [0,1)$, which we use to help us to describe the model. We see here that the quadratic term has a negative sign, and the quartic term is positive, giving rise to spontaneous symmetry breaking, which is in general required for the presence of topological structure. To see how this works, we write the equation of motion as
\be
r^2\frac{d^2\phi}{dr^2}+ r \frac{d\phi}{dr}+\frac{2\phi(1-\phi^2)}{(1-s)^2}=0,
\ee
which can be solved analytically to give
\be\label{phi4}
\phi_{s}(r)=\frac{1-r^{2/(1-s)}}{1+r^{2/(1-s)}}.
\ee
There is another solution, with the minus sign, which behaves similarly. The solution \eqref{phi4} is depicted in Fig.~\ref{fig1} for two distinct values of $s$, for $s=0.6$ and $s=0.8$, and we see that the scalar field varies smoothy from $1$ to $-1$ as $r$ increases in the interval $[0,\infty)$. We note that the solution is sharper for the larger value of $s$. In fact, we have investigated the solution for several values of $s$, and we depicted the cases for $s=0.6$ and $0.8$ to illustrate the general behavior; we shall follow this to depict the figures for $s=0.6$ and for $s=0.8$, until Fig.~\ref{fig12}. 

The corresponding energy density is given by
\be
\rho_{s}(r)=\frac{16 r^{2(1+s)/(1-s)} }{(1-s)^2(1+r^{2/(1-s)})^4 },
\ee
and the total energy is
\be
E=\frac{8\pi}{3(1-s)}.
\ee

This model was investigated before in \cite{s17}. We take it as an example of a skyrmion with skyrmion number $Q=1$. To see this, we note that the solution \eqref{phi4} is such that $\phi(0)=1$, and $\phi(\infty)=-1$. These are the boundary conditions we have used to write such solution, inspired from the behavior of the polynomial \eqref{p4}, which has minima values at $\phi=\pm1$. These two values describe the two degenerate zero energy configurations the system engenders.
In this way, the finite energy topological structure which is described by the solution \eqref{phi4} allows that we describe a skyrmion, with the magnetization as in \eqref{M} and \eqref{T}. We then take $\delta=0$ to get that ${\bf M}(0)={\hat z}$, and ${\bf M}(\infty)=-{\hat z}$. With this, we get that $Q=1$, so we have a skyrmion with skyrmion number $1$.

\begin{figure}[t!]
\centerline{\includegraphics[scale=0.38]{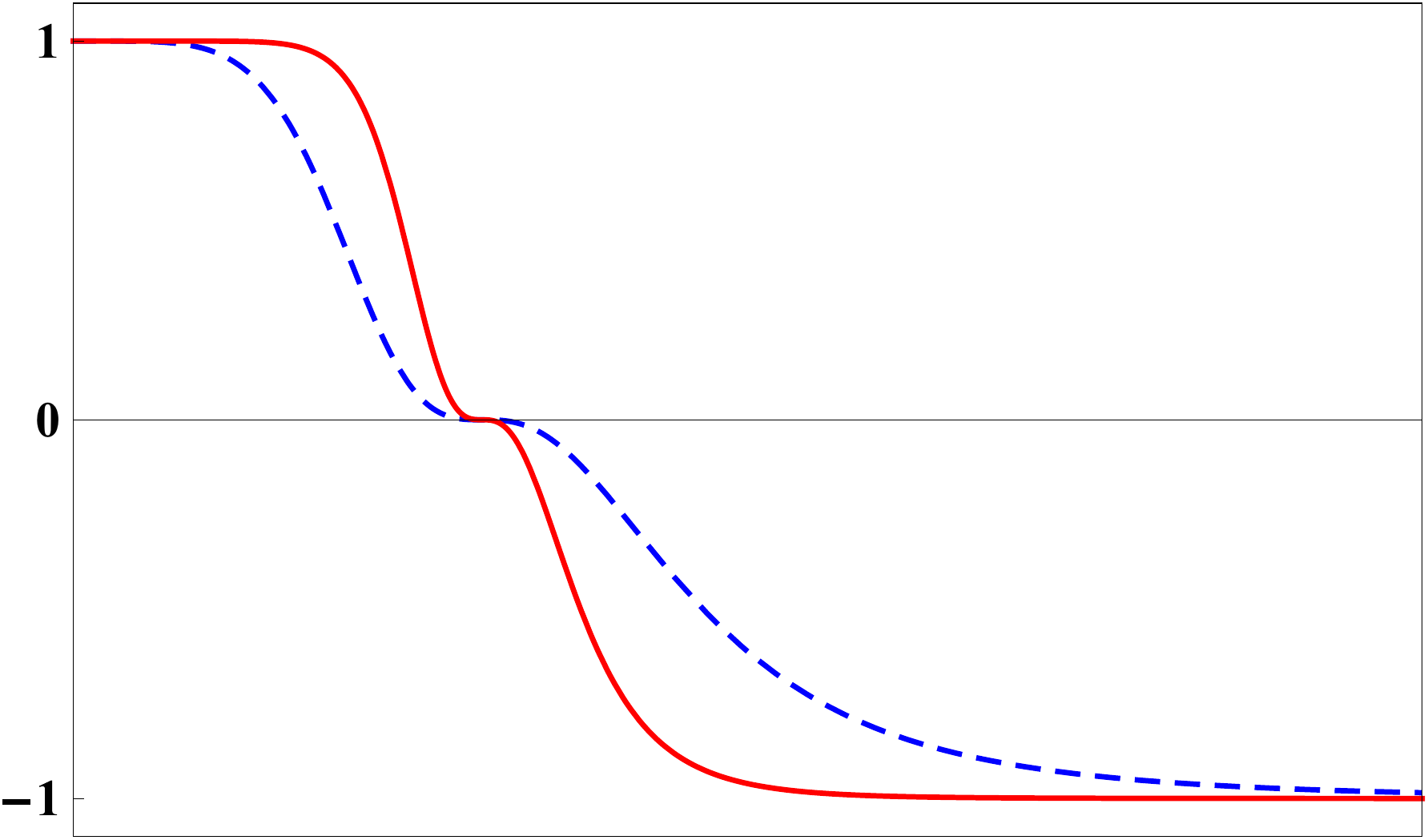}}
\caption{(Color online) The solution $\phi(r)$ for the model \eqref{pma} for $p=3$, depicted for $s=0.6$ and for $s=0.8$, with dashed/blue and solid/red lines, respectively.}\label{fig3}
\end{figure}
\begin{figure}[t!]
\centerline{\includegraphics[scale=0.38]{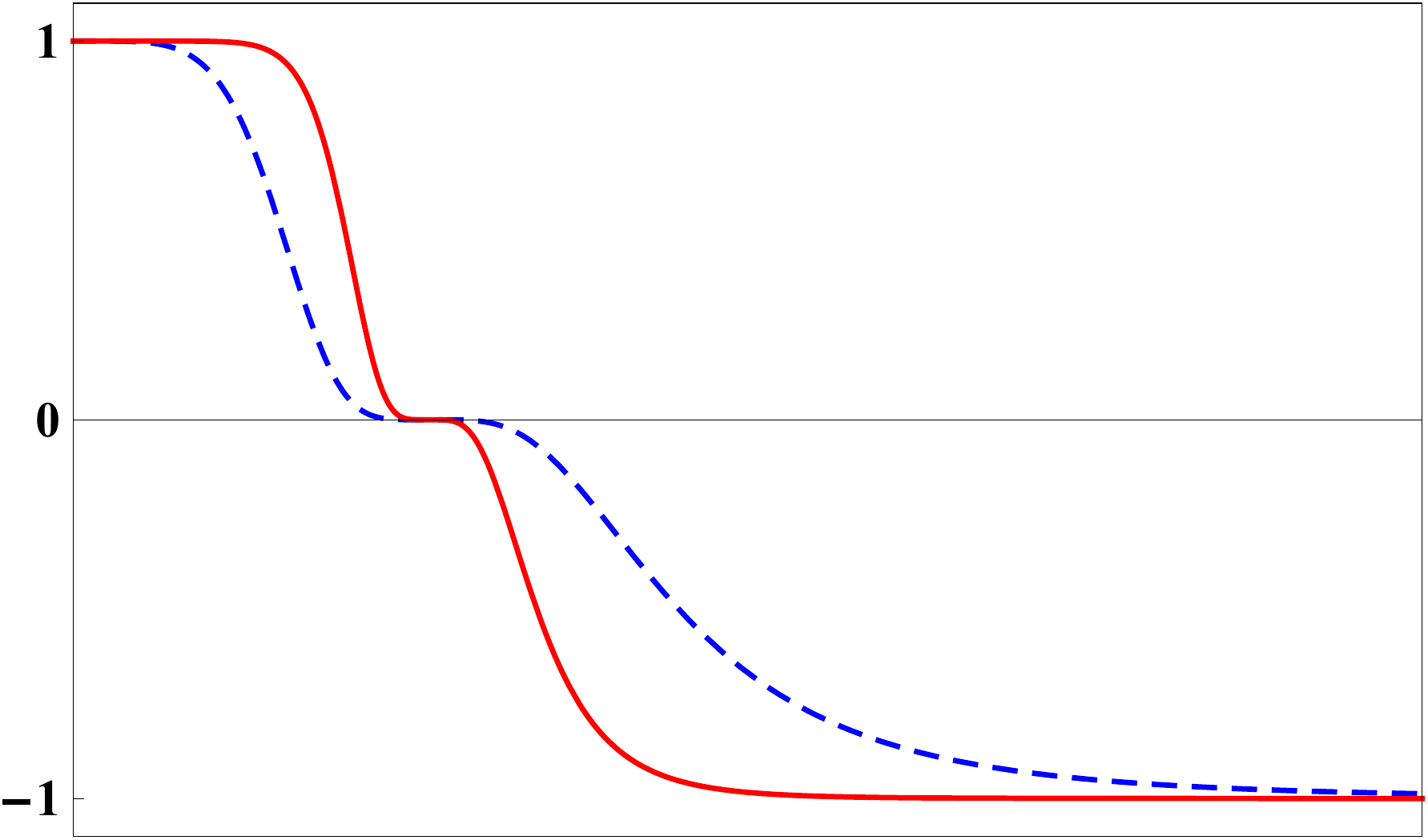}}
\caption{(Color online) The solution $\phi(r)$ for the model \eqref{pma} for $p=5$, depicted for $s=0.6$ and for $s=0.8$, with dashed/blue and solid/red lines, respectively.}\label{fig4}
\end{figure}

\subsection{Vortex} 

Let us now consider another model, with the polynomial having up to the sixth-order power in the field. One takes 
\be\label{p6}
P(\phi)=\frac1{(1-s)^2}\phi^2(1-\phi^2)^2.
\ee
In this case, the quadratic term is positive, the quartic is negative, and the next one is positive, of the sixth-order power in the field. As in the previous case, it gives rise to spontaneous symmetry breaking, but now it also supports a symmetric phase at the origin. To see how this works, we write the equation of motion as
\be
r^2\frac{d^2\phi}{dr^2}+ r \frac{d\phi}{dr}-\frac{\phi(1-\phi^2)^2}{(1-s)^2}+\frac{2\phi^3(1-\phi^2)}{(1-s)^2}=0,
\ee
and the solution is
\be\label{phi6}
\phi_s(r)=\frac{r^{1/(1-s)}}{\sqrt{1+r^{2/(1-s)}}}.
\ee
It is depicted in Fig.~\ref{fig2} for two distinct values of $s$, and here we also note that the solution is sharper for the larger value of $s$. The energy density has the form
\be
\rho_s(r)= \frac{r^{2/(1-s)}}{(1-s)^2(1+r^{2/(1-s)})^{3}},
\ee
and the total energy is given by
\be
E= \frac{\pi^2 s}{2  \sin{(\pi s)}}.
\ee

This model was also studied in \cite{s17}, and it represents a vortex, or a skyrmion with skyrmion number $1/2$, as we now explain. We note that the solution \eqref{phi6} is such that $\phi(0)=0$ and $\phi(\infty)=1$. This is similar to the previous case, and so we follow the same steps, using \eqref{M} and \eqref{T} and taking $\delta=\pi/2$, to see that the magnetization obeys ${\bf M}(0)={\hat z}$, and ${\bf M}(\infty)={\hat\theta}$. We use this to get that $Q=1/2$, so we have a vortex, a skyrmion with skyrmion number $1/2$.

\begin{figure}[t!]
\centerline{\includegraphics[scale=0.38]{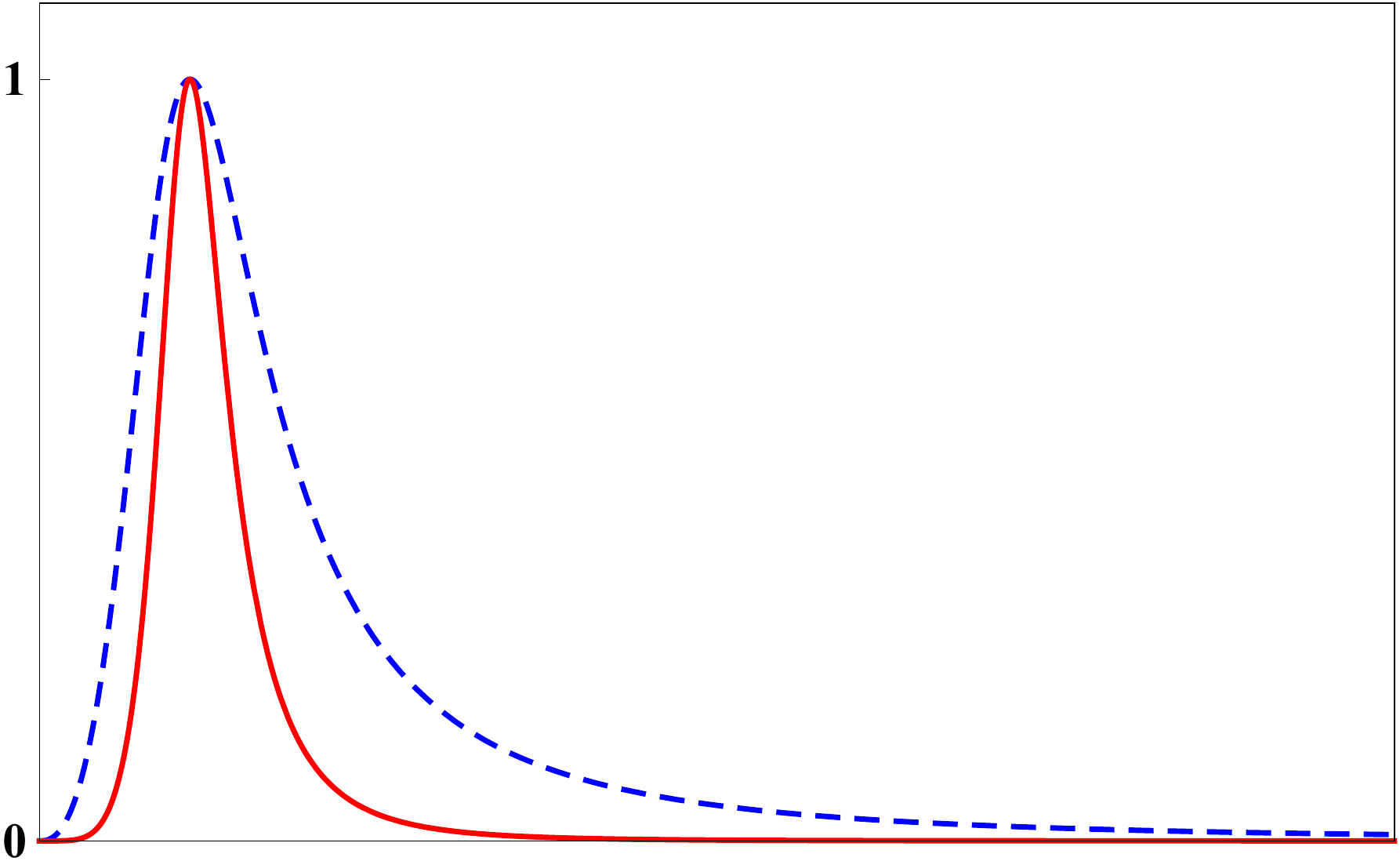}}
\caption{(Color online) The solution $\phi(r)$ for the model \eqref{p4i} with $n=1$, depicted for $s=0.6$ and for $s=0.8$, with dashed/blue and solid/red lines, respectively.}\label{fig5}
\end{figure}
\begin{figure}[t!]
\centerline{\includegraphics[scale=0.38]{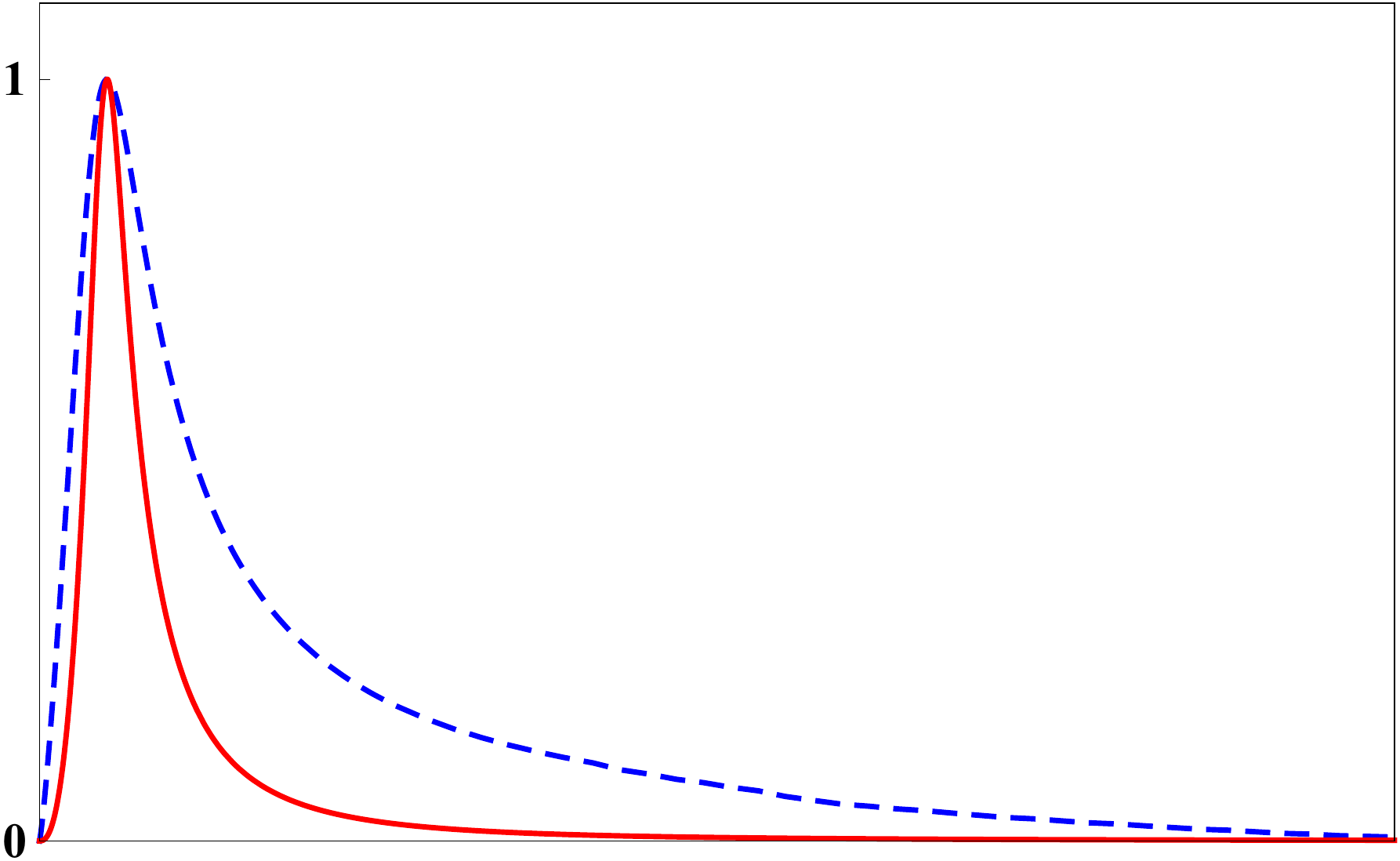}}
\caption{(Color online) The solution $\phi(r)$ for the model \eqref{p4i} with $n=2$, depicted for $s=0.6$ and for $s=0.8$, with dashed/blue and solid/red lines, respectively.}\label{fig6}
\end{figure}

\section{New structures}
\label{nt}

We now search for other models, which present topological profile with skyrmion number $1$. One gets inspiration from a model introduced in \cite{bmm} to study the polynomial
\be\label{pma}
P_{p}(\phi)=\frac{p^2}{(1-s)^2}\phi^2 (\phi^{-1/p}-\phi^{1/p})^2,
\ee
where $p=1,3,5,\cdots,$ is odd integer. One notes that for $p=1$ it gets back to the first model, given by \eqref{p4}. In the general case, the equation of motion has the form
\begin{eqnarray}
&&r^2\frac{d^2\phi}{dr^2} + r \frac{d\phi}{dr} + \frac{2p^2\phi}{(1-s)^2}+\nonumber \\ &-& \frac{p(p-1)\phi^{(p-2)/p}}{(1-s)^2} 
-\frac{p(p+1)\phi^{(p+2)/p}}{(1-s)^2}=0,
\end{eqnarray}
which can be solved analytically to give
\be\label{psol}
\phi_{p}(r)=\left(\frac{1-r^{2/(1-s)} }{1+r^{2(1-s)}}\right)^p.
\ee
The energy density has the form
\be  
\rho_{p}(r)=\frac{16p^2 r^{2(1+s)/(1-s)}(1-r^{2/(1-s)})^{2p-2}}{ (1-s)^2(1+r^{2/(1-s)})^{2p+2} },
\ee
and the energy is
\be
E_p=\frac{8\pi p^2}{ (4p^2-1)(1-s) }.
\ee
We define the (dimensionless) ratio $\varepsilon_p=E_p/E_1$ to get 
\be
\varepsilon_p=\frac{3p^2}{(4p^2-1)},
\ee 
which does not depend on $s$, and decreases from $1$ to $0.75$ as $p$ increases to larger and larger odd integers. One notes that the energy decreases slowly as $p$ increases to larger values, as it is illustrated in the Table 1 below, where we depict $\varepsilon_p$ for several values of $p$.

\begin{table}[h!]
\centering
     \small
     \caption[]{}
\begin{tabular}{lrrrr}
\hline
\hline
\begin{minipage}[t]{.07\textwidth}\begin{flushleft} $\;\;p=1$\end{flushleft}\end{minipage}&
\begin{minipage}[t]{.09\textwidth}\begin{flushright} $p=3\;\;$\end{flushright}\end{minipage}&
\begin{minipage}[t]{.09\textwidth}\begin{flushright} $p=5\;\;$\end{flushright}\end{minipage}&
\begin{minipage}[t]{.09\textwidth}\begin{flushright} $p=7\;\;$\end{flushright}\end{minipage}\\
\hline
$\;\;1.000$ &$0.771\;\;$	&$0.757\;\;$ &$0.753\;\;$  \\
\hline
\hline
\end{tabular}
\label{tab1}
\end{table}
This model is new, and we depict the solution for $p=3$ in Fig.~\ref{fig3}, for two distinct values of $s$. We see from  Fig.~\ref{fig3} that the solution is different from the previous one, depicted in Fig.~\ref{fig1}, which correspond to $p=1$: it has a kind of internal structure when the field $\phi$ changes sign. This is similar to the behavior found before for magnetic domain walls in constrained geometries, at the nanometric scale \cite{is}. We also depict the solution for $p=5$ in Fig.~\ref{fig4}, noting that the internal structure is now wider than it appears for $p=3$. One notes that the thickness of the internal structure depends on $p$ and increases as $p$ is increased. Thus, the energy of the field configuration decreases as one increases the thickness of the internal structure or the transition region. Motivated by the experimental observation related in Ref.~\cite{is}, we believe that skyrmions with such topological behavior may also appear in constrained geometries at the nanometric scale.

We see that the solution \eqref{psol} obeys $\phi(0)=1$ and $\phi(\infty)=-1$, if $p$ is an odd integer, so it can be used to describe a topological structure with unit skyrmion number. To see this, we use \eqref{M} and \eqref{T} and take $\delta=0$ to get ${\bf M}(0)={\hat z}$ and ${\bf M}(\infty)=-{\hat z}$. This leads to the topological charge $Q=1$. Thus, it can be used to represent a skyrmion with skyrmion number $Q=1$, for $p$ odd integer. We conclude that although the topological charge does not depend on $p$, the thickness of the internal structure increases and the energy decreases as one increases $p$.  

To investigate the size of the topological skyrmions, we follow \cite{s17} and use the mean matter radius
\be
{\bar r}=\frac{\int_0^\infty\rho(r)r^2 dr}{\int_0^\infty\rho(r)r dr}.
\ee
One uses it to define the (dimensionless) ratio $R_{p,s}={\bar r}_p/{\bar r}_1$. It has the form
\be  
R_{p,s}=\frac{\int_0^\infty\rho_p(r)r^2 dr}{\int_0^\infty\rho_1(r)r^2 dr}
\frac{\int_0^\infty\rho_1(r)r dr}{\int_0^\infty\rho_p(r)r dr}
\ee
We investigate $R_{p,s}$ for the model \eqref{pma}, but it depends on $p$ and $s$, and leads to awkward expressions; however, we have seen that it increases with $p$ for several values of $s$, but its variation is less expressive for higher values of $s$. We illustrate this in the table II below, for several values of $s$. One notes that as $p$ increases, the size increases and the energy decreases.

\begin{table}[h!]
\centering
     \small
     \caption[]{}
\begin{tabular}{lrrrr}
\hline
\hline
\begin{minipage}[t]{.07\textwidth}\begin{flushleft} $ $\end{flushleft}\end{minipage}&
\begin{minipage}[t]{.09\textwidth}\begin{flushright} $p=1\;\;$\end{flushright}\end{minipage}&
\begin{minipage}[t]{.09\textwidth}\begin{flushright} $p=3\;\;$\end{flushright}\end{minipage}&
\begin{minipage}[t]{.09\textwidth}\begin{flushright} $p=5\;\;$\end{flushright}\end{minipage}&
\begin{minipage}[t]{.09\textwidth}\begin{flushright} $p=7\;\;$ \end{flushright}\end{minipage}\\
\hline
$\;\;s=0.2$ &$1.000\;\;$	&$1.290\;\;$ &$1.501\;\;$ &$1.673\;\;$ \\
\hline
$\;\;s=0.4$ &$1.000\;\;$	&$1.162\;\;$ &$1.275\;\;$ &$1.366\;\;$ \\
\hline
$\;\;s=0.6$ &$1.000\;\;$	&$1.071\;\;$ &$1.121\;\;$ &$1.159\;\;$ \\
\hline
$\;\;s=0.8$ &$1.000\;\;$	&$1.018\;\;$ &$1.029\;\;$ &$1.039\;\;$ \\
\hline
\hline
\end{tabular}
\label{tab1}
\end{table}

We search for the possibility to construct a structure with non-topological profile. The issue here is to describe a localized excitation with zero skyrmion number, as it is suggested in Ref.~\cite{s12x}. To achieve this goal, we have to introduce a polynomial $P(\phi)$ which engenders a single zero-energy minimum, with nonzero values of $\phi$ that make the polynomial vanish. We do this with the choice 
\be\label{p4i}
P_n(\phi)=\frac1{n^2(1-s)^2}\phi^2(1-\phi^{2n}),
\ee
where $n=1,2,3,\cdots$. This is motivated by Ref.~\cite{lump}, and we see that $\phi=0$ is a zero and a minimum of the polynomial. One also notes that there are two other points, $\phi=\pm1$, which are zeros of the polynomial. One sees that the quadratic term is positive, but the other term is negative, such that the spontaneous symmetry breaking phenomenon cannot take place anymore. To see how the system works in this case, we write the equation of motion as
\be
r^2\frac{d^2\phi}{dr^2}+ r \frac{d\phi}{dr}-\frac{\phi(1-(1+n)\phi^{2n})}{n^2(1-s)^2}=0,
\ee
and search for solution with the boundary conditions $\phi(0)=0$ and $\phi(\infty)=0$, to make the solution non-topological. Guided by this, we found the analytical solution
\be\label{phi4inva}
\phi_n(r)=\left(\frac{2r^{1/(1-s)}}{1+r^{2/(1-s)}}\right)^{1/n}.
\ee

The corresponding energy density has the form
\be
\rho_n(r)= \frac{ (2r^{1/(1-s)})^{2/n}(1-r^{2/(1-s)})^{2}}{n^2(1-s)^2r^2(1+r^{2/(1-s)})^{2(n+1)/n}},
\ee
and the total energy is given by
\be
E_n= \frac{\pi^{3/2}\Gamma(1/n)}{n^2(1-s)\Gamma((3n+2)/2n)},
\ee
where $\Gamma(\xi)$ is gamma function. As before, one introduces the ratio $R_n=E_n/E_1$, which has the form
\be
R_n=\frac{\Gamma(1/n)\Gamma(5/2)}{n^2\Gamma(1)\Gamma((3n+2)/2n)}.
\ee
It varies from $1$ to smaller and smaller values, as $n$ increases to larger and larger values.

The solution for $n=1$ is depicted in Fig.~\ref{fig5} for two distinct values of $s$, and we also note that the solution is sharper for the larger value of $s$. The solutions for $n=2$ is depicted in Fig.~\ref{fig6}, and is similar to the case $n=1$.

We now use \eqref{M} and \eqref{T} and take $\delta=\pi/2$ to make the magnetization to obey ${\bf M}(0)={\hat z}$, and also, ${\bf M}(\infty)={\hat z}$. The choice makes the skyrmion number to vanish, so we are dealing with non-topological structures, with $Q=0$. This is a concrete realization of non-topological structures.

\begin{figure}[t!]
\centerline{\includegraphics[scale=0.38]{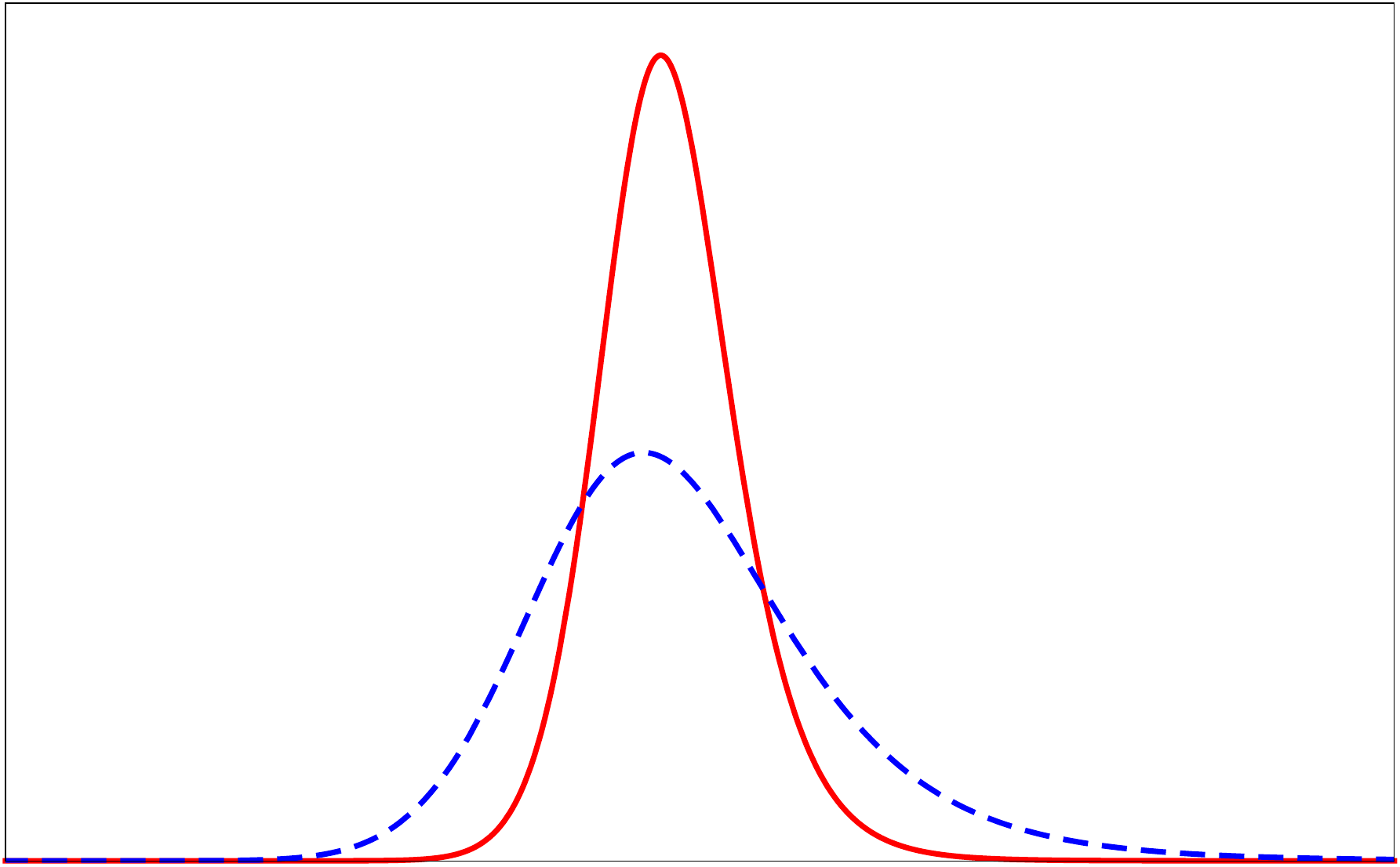}}
\caption{(Color online) The topological charge density \eqref{tcd4} for the model \eqref{p4}, depicted for $s=0.6$ (dashed/blue line) and for $s=0.8$ (solid/red line).}\label{fig7}
\end{figure}
\begin{figure}[t!]
\centerline{\includegraphics[scale=0.38]{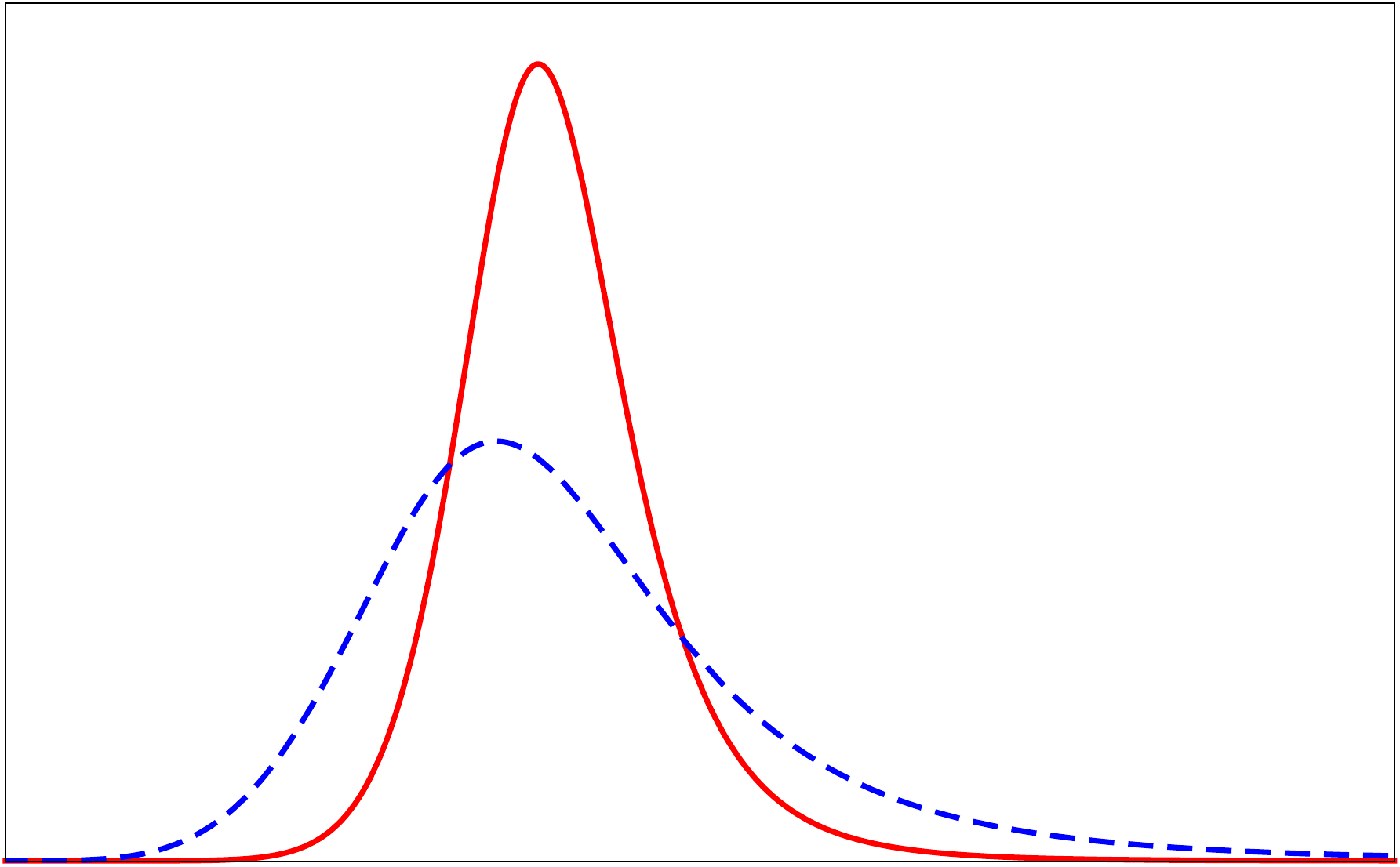}}
\caption{(Color online) The topological charge density \eqref{tcd6} for the model \eqref{p6}, depicted for $s=0.6$ (dashed/blue line) and for $s=0.8$ (solid/red line).}\label{fig8}
\end{figure}

\section{Linear stability}
\label{sta}

Let us now study stability of the solutions found above, against spherically symmetric deformations, using that $\phi=\phi_s(r)+\epsilon\,\eta_s(r)$, with $\epsilon$ being very small real and constant parameter \cite{s17}. We expand the total energy \eqref{Ea} in terms of $\epsilon$ in \eqref{Eb} in the form
\be
E_\epsilon =E_0+\epsilon E_1+\epsilon^2 E_2+\cdots
\ee
where $E_i, i=0,1,2,...$, is the contribution to the energy at order $i$ in $\epsilon$. For the model \eqref{p4}, $E_i$ goes up to $4$, for the model \eqref{p6}, up to $6$; and so on. Of course, $E_0$ is the energy of the solution $\phi_s(r)$, and $E_1$ must be zero. For the model \eqref{p4}, the zero mode has the form
\be 
\eta(r)=A_s\frac{r^{2/(1-s)}}{(1+r^{2/(1-s)})^2},
\ee
where $A_s$ is normalization constant. We can prove that $E_2=0$, $E_3=0$, and $E_4$ has the form
\be 
E_4= \frac{3\pi}{35(1-s)}.
\ee
One sees that it is positive, $E_4>0$, which shows that the solution $\phi_s(r)$ given by \eqref{phi4} is stable against spherically symmetric fluctuations.

\begin{figure}[t!]
\centerline{\includegraphics[scale=0.38]{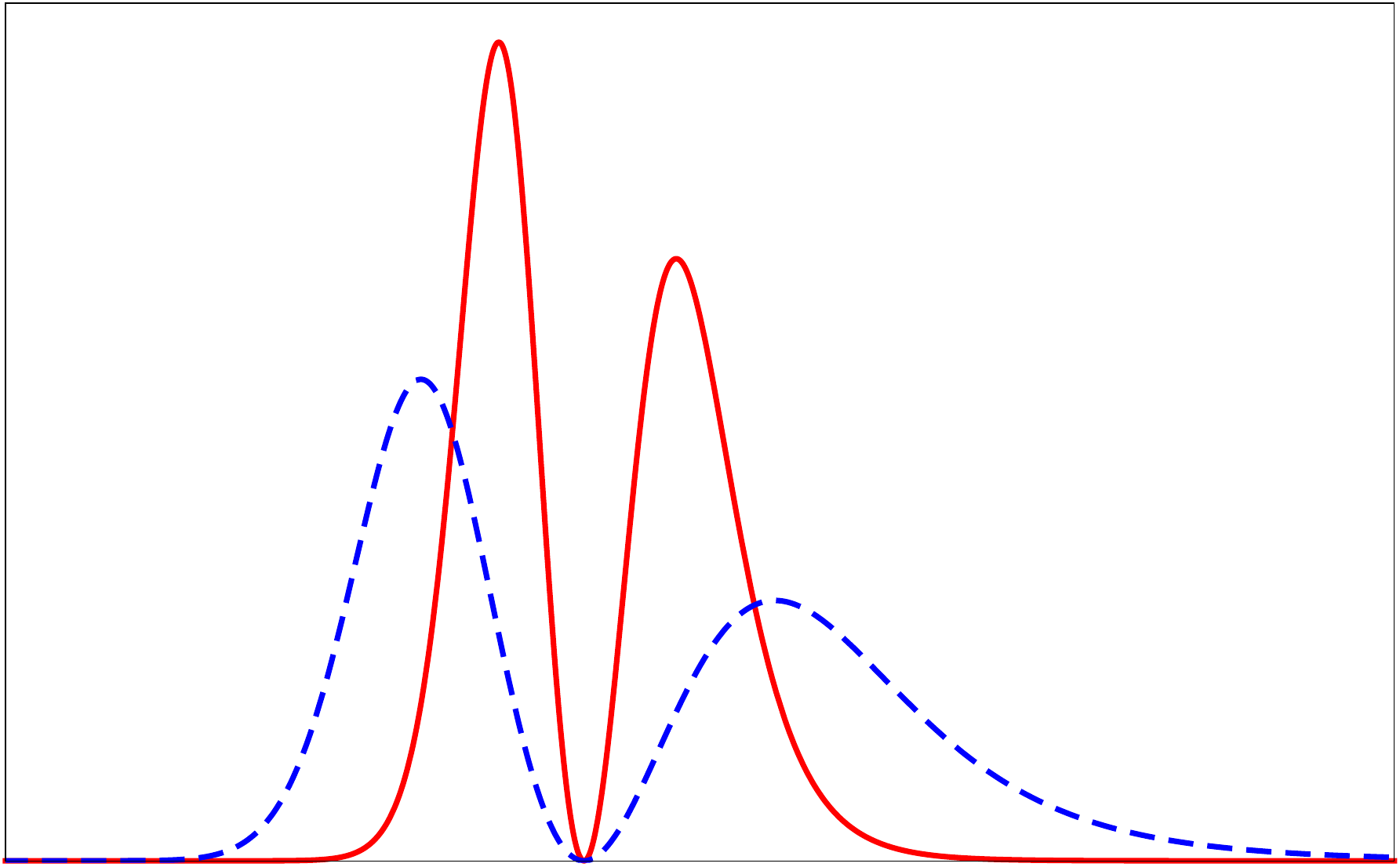}}
\caption{(Color online) The topological charge density \eqref{tcd3ma} for the model \eqref{pma}, with $p=3$, depicted for $s=0.6$ (dashed/blue line) and for $s=0.8$ (solid/red line).}\label{fig9}
\end{figure}
\begin{figure}[t!]
\centerline{\includegraphics[scale=0.38]{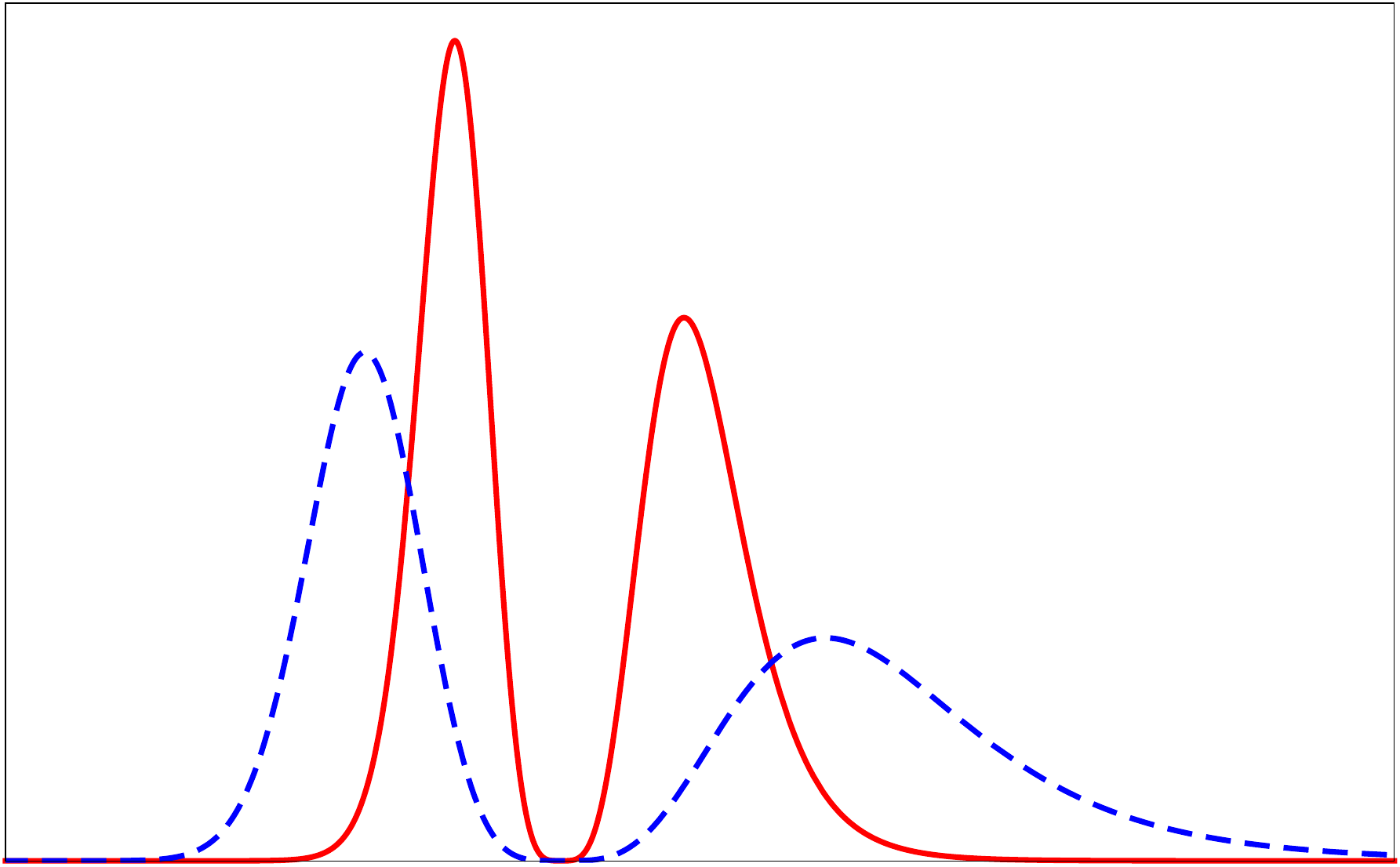}}
\caption{(Color online) The topological charge density \eqref{tcd3ma} for the model \eqref{pma}, with $p=5$, depicted for $s=0.6$ (dashed/blue line) and for $s=0.8$ (solid/red line).}\label{fig10}
\end{figure}

For the model \eqref{p6}, one gets the zero mode
\be  
\eta_s(r)=A_s\frac{r^{2/(1-s)}}{(1+r^{2/(1-s)})^{3/2}},
\ee
where $A_s$ is normalization constant. We can show that $E_2=0$ and
\be
E_3=\frac{15\pi^ 2}{64(1-s)}.
\ee
One sees that it is positive, $E_3>0$, and this shows that the spherically symmetric solution \eqref{phi6} is stable against spherically symmetric fluctuations.

For the model \eqref{pma}, we get the zero mode
\be 
\eta_s^p(r)=A_s^p\frac{r^{2/(1-s)}(1-r^{2/(1-s)})^{p-1}}{(1+r^{2/(1-s)})^{p+1}},
\ee
where $A_s^p$ is normalization constant. The model is controlled by the two parameters $p$ and $s$, and the results show that
$E_2=0$, $E_3=0$, and 
\be 
E^{p,s}_4=\frac{9(p^2-4)(4p^2-5)\pi}{p^2(64p^6-560p^4+1036p^2-225)(1-s)}.
\ee
It is positive for $p=1,3,\cdots$, and for $s\in[0,1)$, so the solution is stable against spherically symmetric fluctuations.

\begin{figure}[t!]
\centerline{\includegraphics[scale=0.38]{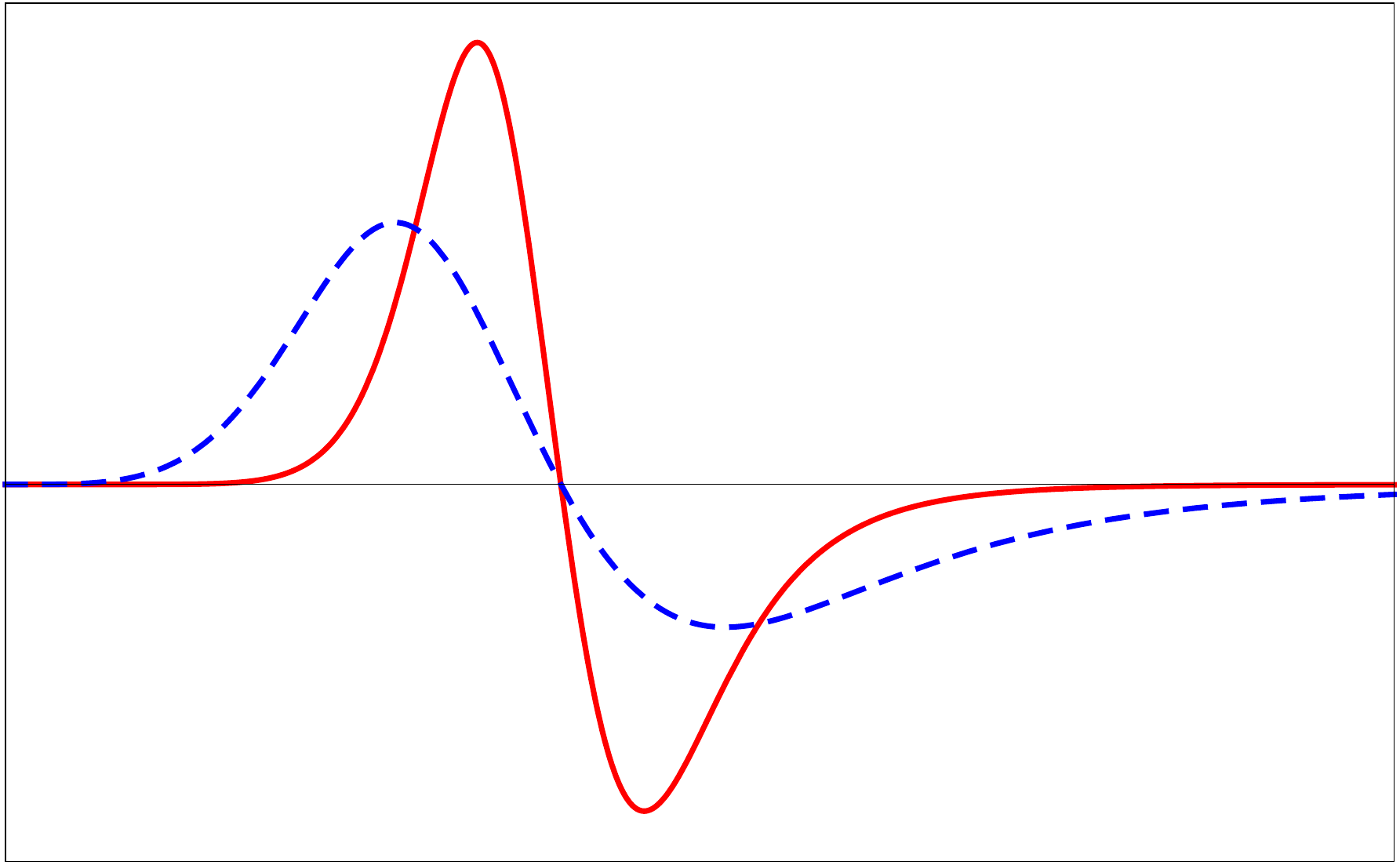}}
\caption{(Color online) The topological charge density \eqref{tcd4ia} for the model \eqref{p4i} with $n=1$, depicted for $s=0.6$ (dashed/blue line) and for $s=0.8$ (solid/red line).}\label{fig11}
\end{figure}
\begin{figure}[t!]
\centerline{\includegraphics[scale=0.38]{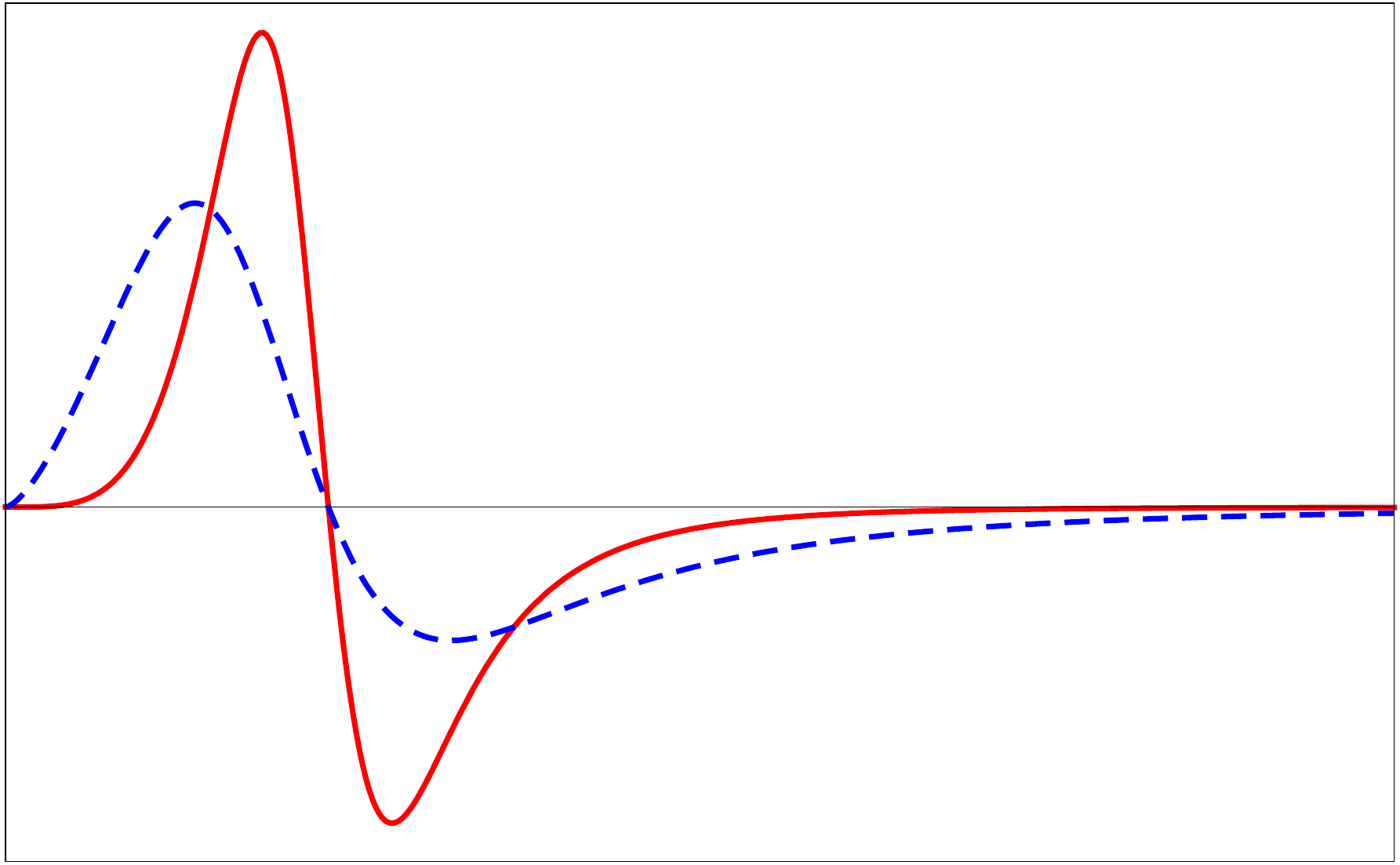}}
\caption{(Color online) The topological charge density \eqref{tcd4ia} for the model \eqref{p4i} with $n=2$, depicted for $s=0.6$ (dashed/blue line) and for $s=0.8$ (solid/red line).}\label{fig12}
\end{figure}

For the model \eqref{p4i}, the zero mode has the form
\be  
\eta_s^n(r)=A_s^n\frac{r^{1/n(1-s)}(1-r^{2/(1-s)})}{(1+r^{2/(1-s)})^{(n+1)/n}},
\ee
where $A_s^n$ is normalization constant. This model is controlled by the two parameters $n$ and $s$, and we could verify that $E_2=0$, $E_3=0$, and that
\be  
E^{n,s}_4=-\frac{2^{(n-2)/n}\pi(2n-1)(10n^2-13n+9)}{3n(1-s)\Gamma(6+2/n)}F_n(\Gamma)
\ee
for
\ben 
F_n(\Gamma) &=& 3(\Gamma(3+{1}/{n}))^2+\Gamma(1+{1}/{n})\Gamma(5+{1}/{n})+\nonumber\\
&&-4\Gamma(2+{1}/{n})\Gamma(4+{1}/{n}).
\een
It is negative for $n=1,2,\cdots$ and for $s\in[0,1)$, and this proves that the spherically symmetric solutions are unstable against spherically symmetric fluctuations. The non-topological structures are neither protected by topology nor against spherically symmetric fluctuations. 

Some of the above results describe distinct topological solutions, with skyrmion number $1$, and a topological vortex, with skyrmion number $1/2$. The skyrmions and vortex are stable against small radial fluctuations. We have also found several distinct non-topological solutions; they have vanishing skyrmion number and are unstable against small radial fluctuations. The results illustrate that the topological structures are protected by topology, while the non-topological ones are not protected.

\section{Topological charge density}
\label{td}

We now deal with the topological behavior of the several distinct solutions described in the previous sections. One uses the topological charge \eqref{Q} to define the topological charge density or the skyrmion number density as follows: take  the topological charge in the form
\be\label{tcd}
Q=\int_0^\infty dr \,q,
\ee
such that the skyrmion number density \cite{s14x} can be written as
\be\label{chi}
q= \frac{r}{2} \,{\bf M}\cdot\partial_x{\bf M}\times\partial_y{\bf M}.
\ee
Now, use  \eqref{M} and \eqref{T} to write
\be\label{tcd1}
q(r)=-\frac{\pi}{4}\cos{\left(\frac{\pi}{2}\phi(r)+\delta\right)}\frac{\partial\phi(r)}{\partial r}.
\ee

We apply this procedure to the models investigated above. 
For the model \eqref{p4} one uses $\delta=0$ to get
\be\label{tcd4}
q(r)=q_0(r)\cos{\frac{\pi}{2}\left(\frac{1-r^{2/(1-s)}}{1+r^{2/(1-s)}}\right)},
\ee
where 
\be 
\displaystyle q_0(r) = \frac{\pi r^{(1+s)/(1-s)}}{(1-s)(1+r^{2/(1-s)})^2}.
\ee
which is depicted in Fig.~\ref{fig7} for two distinct values of $s$. For the model \eqref{p6}, one takes $\delta=\pi/2$ to get
\be\label{tcd6}
q(r)=q_0(r) \sin\frac{\pi}{2}\left(\frac{r^{1/(1-s)}}{(1+r^{2/(1-s)})^{1/2}}\right),
\ee
where
\be 
\displaystyle q_0(r) = \frac{\pi r^{s/(1-s)}}{4(1-s)(1+r^{2/(1-s)})^{3/2}},
\ee
which is depicted in Fig.~\ref{fig8} for two values of $s$. The two Figs.~\ref{fig7} and \ref{fig8} show that the skyrmion numbers behave similarly. Moreover, for the model \eqref{pma} one takes $\delta=0$ to get
\be\label{tcd3ma}
q(r)=q_0(r)\cos{\frac{\pi}{2}\left(\frac{1-r^{2/(1-s)} }{1+r^{2(1-s)}}\right)^p},
\ee
where 
\be 
\displaystyle q_0(r)=\frac{p\pi r^{(1+s)/(1-s)}(1-r^{2/(1-s)})^{p-1}}{(1-s)(1+r^{2/(1-s)})^{p+1}}. 
\ee
We depict in Fig.~\ref{fig9} the case with $p=3$, and in Fig.~\ref{fig10} the case with $p=5$, for two values of $s$. The two figures show very clearly the presence of the two-peak profile in the skyrmion number density, and this is different from the two previous cases. In the models that describe skyrmions with skyrmion number $Q=1$, in the first model, the skyrmion number density has a single peak, while the others have the two-peak profile. This behavior shows that the skyrmions for $p=3,5,\cdots,$ are different, and we further comment on this in the next section, where we deal with the topology of the above solutions.

For the non-topological structures we take $\delta=\pi/2$. 
The model \eqref{p4i} gives
\be\label{tcd4ia}
q(r)=q_0(r)\sin\frac{\pi}{2}
{\left(\frac{2 r^{1/(1-s)}}{1+r^{2/(1-s)}}\right)^{1/n}},
\ee
where
\be 
\displaystyle q_0(r) = \frac{\pi (1-r^{2/(1-s)})}{4n(1-s)(1+r^{2/(1-s)})r}\left(\frac{2r^{1/(1-s)}}{1+r^{2/(1-s)}}\right)^{1/n}. 
\ee

We depict in Fig.~\ref{fig11} the case with $n=1$, and in Fig.~\ref{fig12} the case with $n=2$, for two distinct values of $s$. We see that the skyrmion number densities of the non-topological structures are similar to each other, but they are well different from the behavior presented by the topological structures. They change sign as $r$ increases to larger values, and this is required to make the topological charge vanish.

\section{Topology}
\label{top}

\begin{figure}[t!]
\centerline{\includegraphics[scale=0.166]{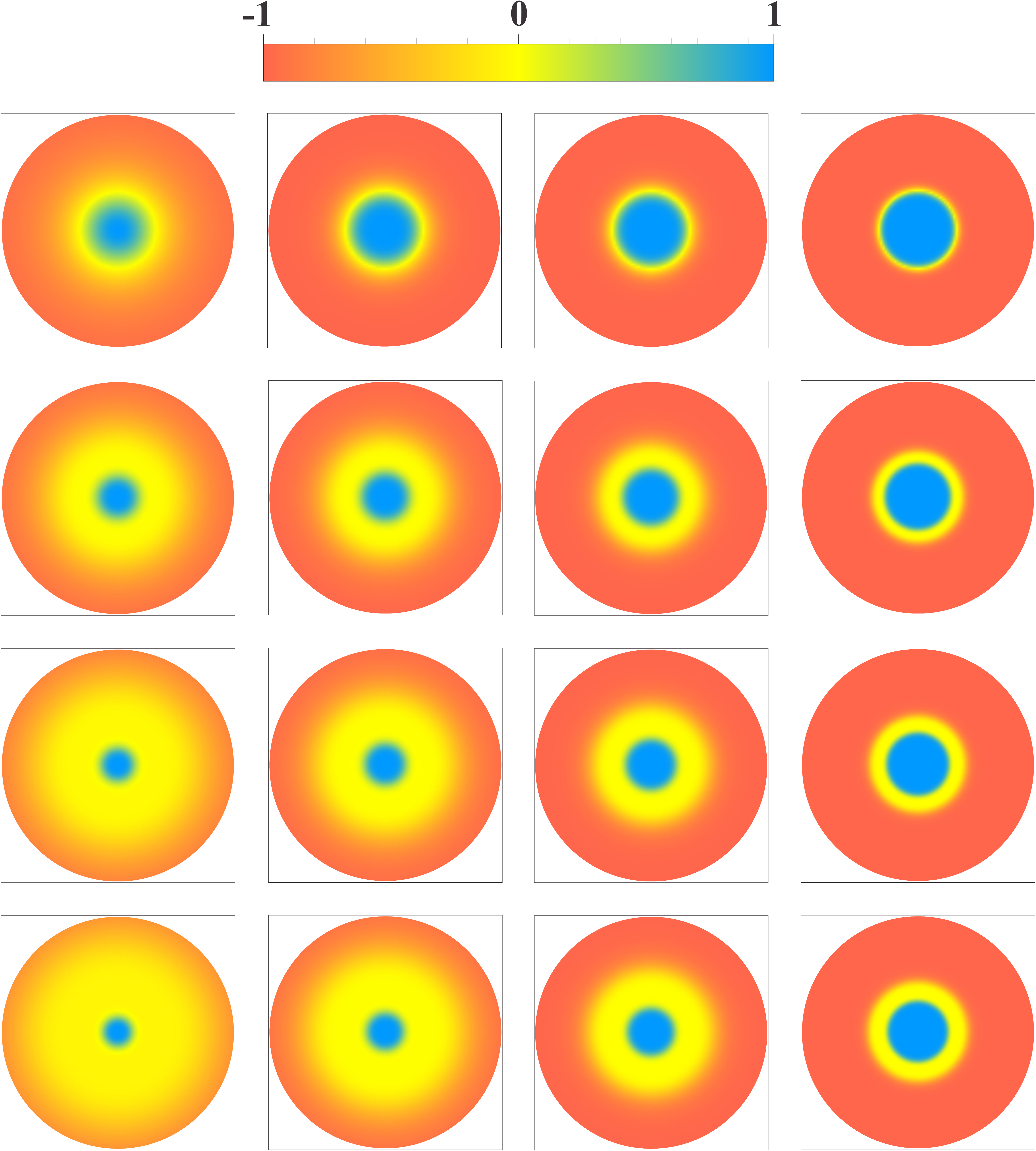}}
\caption{(Color online) The topological structure with skyrmion number $1$, which is controlled by the model \eqref{pma}, depicted for $p=1,3,5,7$ (from top to bottom) and for $s=0.2, 0.4, 0.6, 0.8$ (from left to right).}\label{fig13}
\end{figure}

In order to highlight the internal structure or transition region that appears in the skyrmions with skyrmion number $Q=1$, we follow \cite{s13,s14x} and use a continuum of colors, taking red to represent the magnetization pointing downward in the ${\hat z}$ direction, yellow for the magnetization vanishing along the  ${\hat z}$ direction, and blue for the magnetization pointing upward in the ${\hat z}$ direction. We depict these magnetic structures in Fig.~\ref{fig13}, for $p=1, 3, 5, 7$ and for $s=0.2, 0.4, 0.6, 0.8$, to illustrate how the skyrmion changes as we vary $p$ and $s$. These profiles follow the solutions themselves, and one notes that the larger the value of $s$ is, the sharper the solution becomes.

As displayed in Fig.~\ref{fig13}, the skyrmions of the models with $p=3, 5, 7$ have larger and larger transition regions, depicted in yellow, when compared to the skyrmion of the first model, with $p=1$. This behavior is similar to the effect investigated before in \cite{s14x}, where a larger transition region has been identified, correlated with the two-peak profile which appeared in the skyrmion number density there studied. The effect seen in \cite{s14x} is due to the Rashba spin-orbit coupling. In the current work we are using another route, proposed in \cite{s17}, and here we could also map internal or transition regions, adding fractional power to the scalar field. It is interesting to remark that scalar field with such fractional power was first suggested in \cite{bmm}, and in the one-dimensional case, the domain wall it generates correctly maps structures found in \cite{is}, in the magnetic material ${\rm Fe}_{20}{\rm Ni}_{80}$ in constrained geometries. The results suggest that the current procedure is of direct interest to the subject, since it provides an analytical route to describe skyrmion features, capable of inducing internal structure to such spin textures. It is worth mentioning here that the internal structure of a skyrmion has many degrees of freedom which could be modified. To give an example, we quote the recent work \cite{ezawa} where a high-topological-number skyrmion with complex internal structure is obtained in a nano-contact device.

\begin{figure}[t!]
\centerline{\includegraphics[scale=0.25]{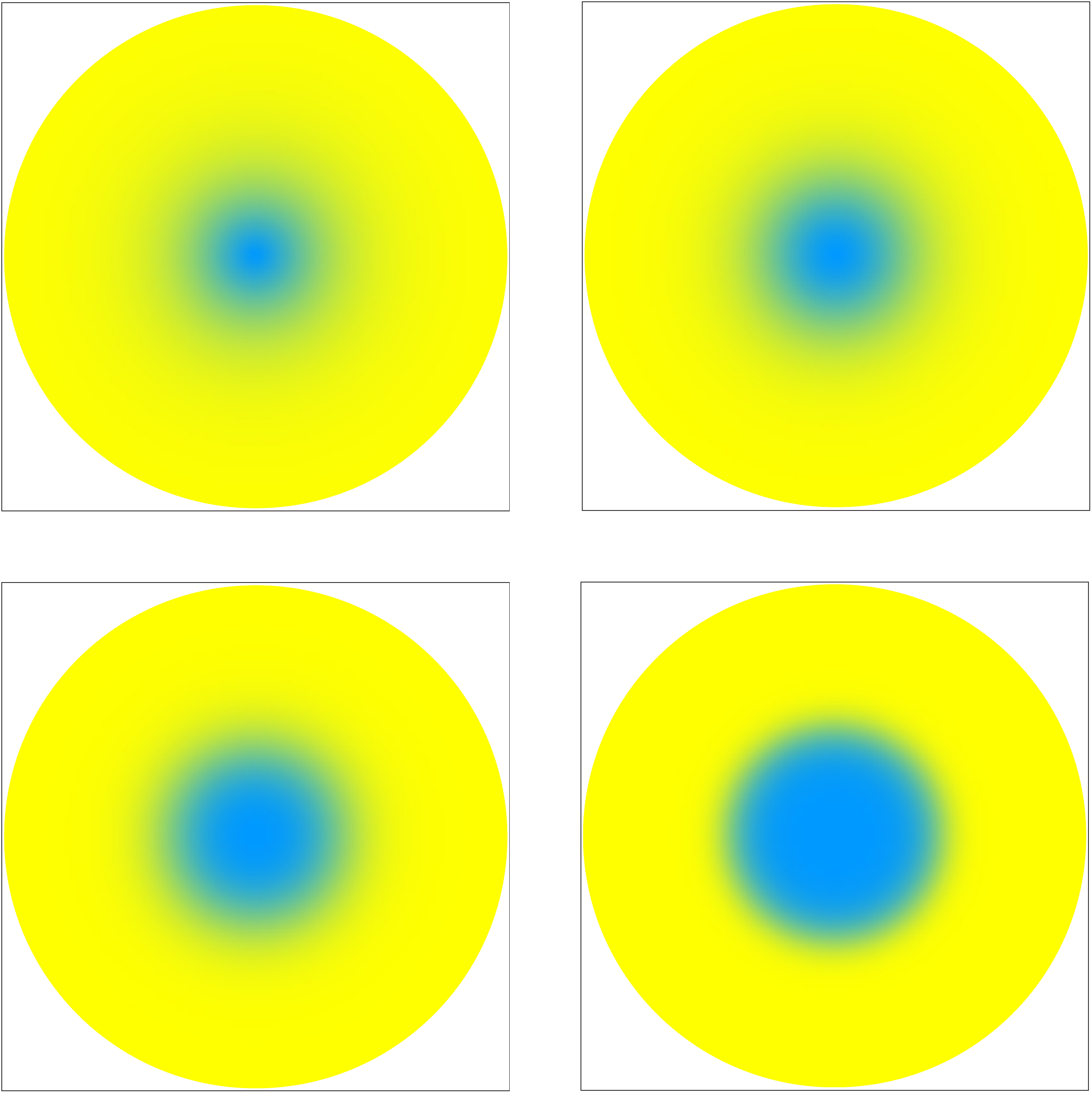}}
\caption{(Color online) The topological structure with skyrmion number $1/2$ which is controlled by the model \eqref{p6}, depicted for $s=0.2, 0.4, 0.6, 0.8$, from top left to bottom right.}\label{fig14}
\end{figure}

\begin{figure}[t!]
\centerline{\includegraphics[scale=0.24]{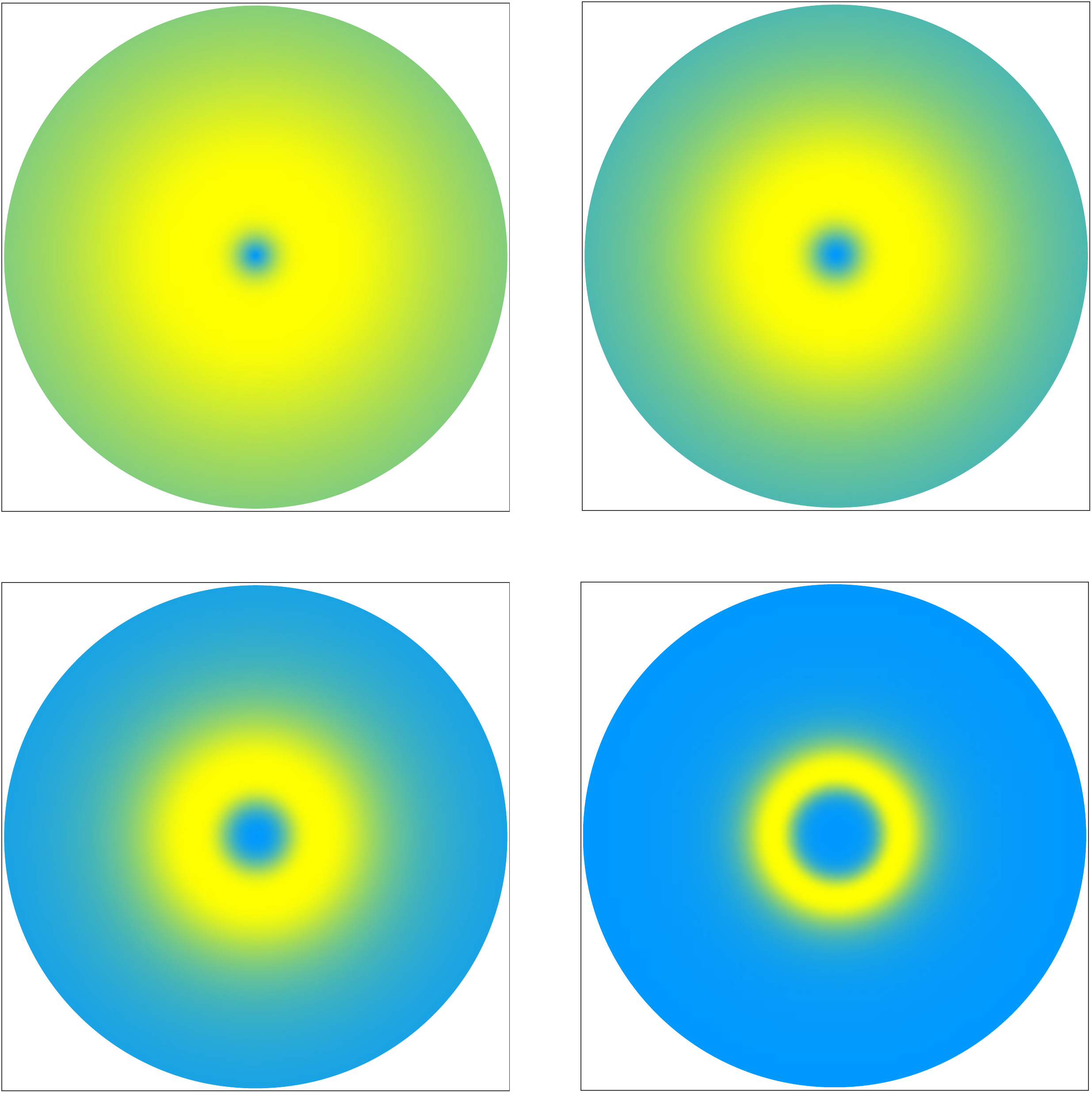}}
\caption{(Color online) The non-topological structure with vanishing skyrmion number, controlled by the model \eqref{p4i}, depicted for $n=1$, and for $s=0.2, 0.4, 0.6, 0.8$, from top left to bottom right.}\label{fig15}
\end{figure}

The procedure is capable of describing vortices, with skyrmion number $Q=1/2$, and non-topological structures, with vanishing skyrmion number. In Fig.~\ref{fig14} we illustrate the vortex of the model \eqref{p6}, and in Fig.~\ref{fig15} the case of vanishing skyrmion number of the model \eqref{p4i} for $n=1$; the colors follow the pattern used in Fig.~\ref{fig13}. The two figures show that the solutions are sharper for larger values of $s$, as it also appeared in the previous cases.

The several solutions constructed in Secs.~\ref{remarks} and \ref{nt} have distinct profiles, which appear from the different ways the polynomials are constructed, and this is deeply connected to the physical behavior of the models. As we see, in the model \eqref{p4} that supports skyrmion with unit skyrmion number, for instance, the $\phi^2$ term is negative, and requires that we add the $\phi^4$ term with positive sign, to stabilize the model. This positive sign means that the fourth-order self-interactions are attractive and stabilize the system. In the second model, which supports vortex with half-integer skyrmion number, we have changed the signs of both the $\phi^2$ and $\phi^4$ terms, but we added a $\phi^6$ term, attractive, which works to stabilize the new model. In the next model, although the $\phi^2$ is positive, there are other fractional powers in the field, and this makes the model well different from the two previous ones. In the last model, the $\phi^2$ term is positive but the term with $\phi^{2n+2}$ is negative, so it makes the self-interactions repulsive and contributes to make the non-topological structure unstable. 

Another interesting result shows that the physical properties of the system are mandatory to describe the behavior of the topological or
non-topological structure. With this in mind, we note that in the work \cite{s12x}, two distinct scenarios are built, with similar vortex structures with skyrmion number $1/2$: one, with a vortex on top of a magnetic material with the magnetization pointing downward, and the other, with a similar vortex on top of the same material, but now with the magnetization pointing upward. Since the magnetization at the center of the two vortices points upward, it seems that the vortex of the first arrangement is a topological structure with skyrmion number $1$, similar to the structure we have constructed from the model \eqref{p4}, and also, that the vortex of the second arrangement is a non-topological structure, similar to the structures we have constructed from the model \eqref{p4i}. They then apply an in-plane magnetic field, which they increases until the vortex is destroyed. 

The experimental data show that the vortex of the first arrangement, which appears to be a topological structure with skyrmion number $1$, is destroyed after the second, requiring a magnetic field which is stronger than the magnetic field necessary to destroy the vortex of the second arrangement. The authors suggest that the topology is playing its game, making the topological structure harder to be destroyed. However, from the above results we believe that another interpretation is possible, because they are using vortex on top of the background material, and that makes the two solutions hybrid configurations, and not genuine skyrmions, as the ones constructed in this work. In particular, we note that the
non-topological structure is unstable, and could not survive the presence of the external, in-plane magnetic field. We believe that it is the background out-of-plane magnetic field that appears from the magnetic material where the vortex stands that acts distinctly, making the vortex in the first arrangement more stable than the other.

To decide on this, we would suggest to modify the background material, letting it be an inert material, supporting no net spin arrangement. With this, we repeat the experiment, but now with the vortex on top of the inert material. We think that the vortex will be destroyed with an in-plane magnetic field with value in between the two values obtained in Ref.~\cite{s12x}. If this is true, we are then measuring how the out-of-plane magnetic field of the background material contributes to weaken or strengthen the topological structure. The result would allow that we manipulate the topology of vortex, an issue which is of current interest to taylor such nanometric spin textures. Evidently, the suggestion requires further dedicated investigations, similar to the one presented in \cite{s12x}, but this is outside the scope of the current work.

Another suggestion we could make is to prepare another experiment, similar to the one done in \cite{s12x}, but now using skyrmions with distinct but larger transition regions, on top of some inert background material. We then turn on and increase the in-plane magnetic field, to measure the topology strength as a function of the width of the transition region. This would bring further light on the topological strength of magnetic skyrmions, with distinct internal structures.

\section{Comments and conclusions}
\label{end}

In this work we studied the topological behavior of localized magnetic structures with skyrmion number $1$, $1/2$, and $0$. They appear from distinct models, with attractive and/or repulsive self-interactions. The two first cases, with skyrmion number $1$ and $1/2$ were studied before in \cite{s17}, the third one, with unit skyrmion number and the two case of vanishing skyrmion number are new and appear here for the first time. The several cases of non-vanishing skyrmion number are protected by topology, and are stable against radial fluctuations, but the cases of zero skyrmion numbers which are non-topological, are linearly unstable, as we investigated explicitly in this work.

As shown in the investigation, the several distinct structures appear from different models, and are controlled by specific self-interactions, which are directly connected with the physical behavior of the models that support the localized structures. For instance, the topological structure with unit skyrmion number that appeared from model \eqref{p4} is formed under the presence of attractive quartic self-interactions, while the non-topological structures with vanishing skyrmion numbers that appeared from model \eqref{p4i} require repulsive self-interactions of the $\phi^{2n+2}$ $(n=1,2,\cdots)$ type. The topological structures are stable against radially symmetric fluctuations, and the non-topological structures are unstable.

An interesting feature of the present study is that it is implemented analytically, and this helps us to better understand some important physical properties of the spin textures that behave as topological vortices and skyrmions, at the nanometric scale. The results of the work motivate us to use the procedure to taylor the spin structures, controlling the way the magnetization varies, due to quantum \cite{s13} and other \cite{X,Y} effects. In particular, we think that skyrmions with internal structures as the ones reported in this work may appear in magnetic materials in constrained geometries \cite{is}. We also believe that the analytical tools used in the current work may be of interest to understand nontrivial aspects of the magnetic skyrmions with higher topological number recently studied in Ref.~\cite{ezawa}. We shall further report on the subject elsewhere.  

\acknowledgments

This work is partially supported by CNPq, Brazil. DB acknowledges support from projects 455931/2014-3 and 306614/2014-6, JGGSR acknowledges support from projects 308241/2013-4 and 479960/2013-5 and EIBR acknowledges support from project 160019/2013-3.


\end{document}